\documentstyle[preprint,aps,psfig]{revtex}

\newcommand{\be}{\begin{eqnarray}}
\newcommand{\ee}{\end{eqnarray}}
\newcommand{\bc}{\begin{center}}
\newcommand{\ec}{\end{center}}

\begin{document}
\draft
\tighten

\title { How unique is the Asymptotic Normalisation Coefficient (ANC) method?}
\author{J.C. Fernandes\footnote{E-mail:jcff@wotan.ist.utl.pt},
R. Crespo\footnote{E-mail:raquel@wotan.ist.utl.pt},   }
\address{ Departamento de F\'{\i}sica, Instituto Superior T\'ecnico, \\
and Centro Multidisciplinar de Astrof\'{\i}sica (CENTRA) \\
Av Rovisco Pais 1096  Lisboa Codex, Portugal}
\author{ F.M. Nunes\footnote{E-mail:filomena@wotan.ist.utl.pt}  }
\address{ Universidade Fernando Pessoa, \\
Praca 9 de Abril, 4200 Porto, Portugal \\
and Centro Multidisciplinar de Astrof\'{\i}sica (CENTRA) \\
Av Rovisco Pais 1096  Lisboa Codex, Portugal}

\date{\today}

\maketitle

\begin{abstract}
The asymptotic normalisation coefficients (ANC)
for the vertex $^{10}$B $\rightarrow$ $^9$Be + p is
deduced from a set of different proton transfer reactions at
different energies. This set should  ensure the peripheral
character of the reaction and availability of data for the elastic channels.
The problems associated with the characteristics of the data 
and the analysis are discussed.
For a subgroup of the set of available data, the uniqueness property
of the extracted ANC is fulfilled. However, more measurements are needed 
before a definite conclusion can be drawn.

\end{abstract}



\section{Introduction}

Considerable effort, both theoretical and experimental,
has been devoted in the last few years to the analysis
of nuclear capture reactions.
At the energies  relevant for Astrophysics, (p,$\gamma$) or
($\alpha, \gamma$) reactions have very low cross section values due to 
the Coulomb barrier repulsion. Thus, in many cases, the only
access to the low energy region is through model dependent
extrapolations of the higher energy  data.
In addition to this experimental limitation, many reactions
of astrophysical interest involve radioactive beams which cannot be
performed using conventional experimental techniques.
The Coulomb dissociation \cite{Moto} and the asymptotic normalisation
coefficients (ANCs) extracted from transfers \cite{xu} have been put
forward recently as alternative methods to obtain information about
the astrophysical S-factors. As recognised by the physics community,
while very appealing,  these methods
need to be subject to severe tests in order to assess their validity
\cite{seattle}. The aim of this work is to check
upon the validity of the ANC method. 

Given the limited sets of data for peripheral transfer reactions,
the results presented here may not be conclusive.
However, we hope that this work will underline the present
difficulties in validating the method and motivate further measurements.

We firstly describe the ANC method (section II). 
Then, we analyse in detail the
different reactions that will be used in the present work (section III).
Particular attention will be paid to the characteristics of the data.
Finally we present a discussion of the results and conclusions in 
section IV.

\section{A systematic study on proton transfer reactions}

The ANC method for the transfer reaction
\be 
A+a \rightarrow   B + b ~~~~~ (a = b + x, ~~  B = A + x),
\ee
relies on two assumptions.
Firstly, the reaction mechanism used to describe the transfer mechanism 
should give direct information  
of the nuclear overlap integrals  $\langle A | B \rangle$,
$\langle a | b \rangle$.
The differential cross section is given by
\be
\frac{d \sigma}{d \Omega} = 
\frac{  \mu_{\rm i } \, \mu_{\rm f}}
{4 \pi^2 \hbar^4}
\frac{k_b}{k_a} \frac{1}{(2J_A + 1) (2 J_a + 1) } \sum |T_{fi}|^2
~~,
\ee
with $ \mu_{ \rm i } $, \, $ \mu_{  \rm f }$ the reduced 
masses for the initial ($A - a$) and final ($b - B$) 
channels and ${k_a},  \,   {k_b}$  the 
incident and outgoing momenta in the centre-of-mass frame.
The DWBA reaction mechanism has been used to analyse the differential cross
section for the transfer reaction.
The transition amplitude for the transfer reaction process  in the post
form is
\be
T_{fi} = \sum  \langle \Psi_f^{(-)} 
 {\cal I}_{  \scriptsize  A B  } | V_{xb} + V_{bA} - U_{bB} | 
{\cal I}_{  \scriptsize  ab} \Psi_i^{(+)} \rangle ~~.
\label{TDWBA}
\ee
In this equation $\Psi_f^{(-)}$ and $ \Psi_i^{(+)}$ are the distorted waves
in the final and initial  channels respectively.
$ {\cal I}_{ \scriptsize  AB}$ 
and ${\cal I}_{ \scriptsize  ab}$ are the
nuclear overlap integrals  $\langle A|B \rangle$ and  $\langle a|b \rangle$.
The remnant term, $ V_{bA} - U_{bB}$, (where 
$ V_{bA}$ is the interaction between the projectile
core  and the target A and $U_{bB}$ the optical potential for
 the outgoing channel), is usually small and may be neglected
but will be included in our calculations.

\begin{figure}[h!]
\centerline{
	\parbox[t]{4.0in}{
	\psfig{file=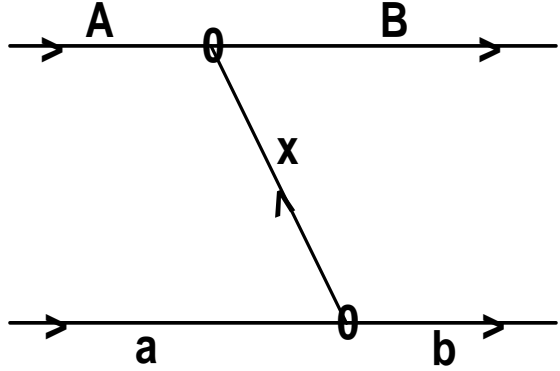,height=3.8in,width=3.8in}}
}
	\caption{Vertex diagram for the transfer reaction A(a,b)B.  }
	\label{fig:vertex}
\end{figure}

Secondly, according to the ANC method, the transfer reaction should be
peripheral so that the asymptotic part of the overlap
integrals gives the dominant contribution to the transition amplitude.
For example, outside the range $R_N$ of the $A$-$x$ nuclear interaction, 
the overlap integral $\langle A | B \rangle$ becomes
\be
{\cal I}_{AB \ell j } \approx C_{AB \ell j}
W_{\eta \ell+ {\scriptstyle\frac{1}{2}} }(2 \kappa r)/r ~~~~~~~~~
r \gg R_N \label{overlap}
\ee
where $W_{\eta \ell+ {\scriptstyle\frac{1}{2}} }(2 \kappa r)$
is the Whittaker function, $\eta = Z_A Z_x e^2 \mu / \kappa$ 
the Sommerfeld parameter, $\kappa = \sqrt{2 \mu \epsilon}/\hbar$,
$\mu$    and  $\epsilon$  the reduced mass and   the binding energy 
for the ($A$--$x$) system.  $C_{AB \ell j}$ is the 
the asymptotic normalisation coefficient (ANC)
for the overlap function  $\langle A | B \rangle$, related to
the asymptotic normalisation of the single particle (s.p.) wave  function
$b_{AC \ell j}$ and a spectroscopic factor ${\cal S}$ by  \cite{us}
\be
C_{AB \ell j} = {\cal S}^{1/2}  \, b_{AB \ell j} \label{ANC-AC}~~.
\ee
The ANC $ C_{AB \ell j}$  defines the vertex for
the virtual transitions $B  \rightarrow A + x$ as shown
in fig.(\ref{fig:vertex}).
It has been shown \cite{xu,us} that as long as the reaction is peripheral,
the ANC is independent of the details of the s.p. parameters
used to describe the nucleus B ground state. That is, the effect of 
different s.p. parameters (which result in different
s.p. asymptotic normalisations)
is compensated by the deduced experimental spectroscopic factors
such that the ANCs become independent of the s.p. model.

For proton transfer reactions, the  extracted ANCs
gives an alternative method of determining the
zero energy  cross section for the capture 
reaction $A + p \rightarrow B + \gamma$ or alternatively  
S$_{1A}(0)$ \cite{xu}, providing of course that the  overlap integral
$\langle a | b \rangle$ is known.
The spectroscopic factor is  obtained (by a $\chi^2$ fit) from
the ratio between the data and the DWBA calculation in the
forward angle region and defined here as ${\cal S}_{\rm exp}$.
To simplify notation we shall omit the angular momenta quantum numbers from 
the ANCs.

By choosing appropriate beam energies and scattering angles such
that the transfer reaction remains peripheral this method is expected
to provide a unique, structure model independent  ANC. 

The ANC method was firstly applied for extracting the S$_{17}$-factor
from the study of the reaction $^{7}$Be(d,n)$^{8}$B \cite{Liu}.
The peripheral character of the reaction and the dependence on the 
optical potential for the incoming and outgoing channels have been 
recently studied  \cite{us}. It was shown  that for the DWBA analysis, the 
optical potentials for the entrance and outgoing channels need to be known
in order to minimise uncertainties on the extracted S-factors \cite{us,comm}.

The ANC method was also applied for extracting the S$_{17}$-factor
from the study of the
$^{10}$B($^{7}$Be,$^{8}$B)$^{9}$Be reaction \cite{Azhari}.
The transfer differential cross section was
measured with high accuracy using an 84 MeV $^7$Be radioactive beam,
in the forward angle region, to ensure its peripheral character. The
optical potentials for the incoming and outgoing channels were derived from
folding model calculations and validated by the elastic data. 
From the  measured transfer differential cross section 
the asymptotic normalisation for the virtual transitions 
$^8$B $\rightarrow ^7$Be + p  and thus the S$_{17}$(0)
 was extracted
assuming that the ANC for the $^{10}$B $ \rightarrow  ^9$Be + p
vertex was known. This ANC
was  determined in the same way from the analysis of  
$^9$Be($^{10}$B,$^9$Be)$^{10}$B  at an incident energy of 100 MeV
\cite{Akram}.

Due to the increasing interest on this method it is timely to perform tests,
to ensure its applicability.
A first test of the ANC method was made in \cite{Gagliardi} 
where the proton transfer reaction $^{16}$O($^3$He,d)$^{17}$F was analysed.
It was shown in that work that the deduced  S--factor for the capture
reaction $^{16}$O(p,$\gamma$)$^{17}$F  agreed well with the capture 
data.   In the present work further tests are performed.

Given the ANC  $C_{AB \ell j}$,  the question we address here 
is: {\em how unique is this value, deduced from different proton transfer 
reactions
or from  the same reaction but at 
different energies, assuming that the peripheral character is 
satisfied ?} For the present  analysis   we choose   the case of the  ANC  
for the $^{10}$B $ \rightarrow  ^9$Be + p  vertex here called
C$_{19}$. This choice was motivated by the accurate forward angle data
for the transfer reaction $^9$Be($^{10}$B,$^9$Be)$^{10}$B, measured 
at a laboratory  energy of 100 MeV, together with good knowledge of
the optical potentials \cite{Akram}.
A literature  search was then performed to find proton transfer 
data at  low energy from which independent values for 
C$_{19}$ could be extracted.
In order to reduce the uncertainties associated with the lack of
knowledge of the optical potential for the incoming and outgoing channel,
our search  was restricted to cases where  elastic scattering data
was available, whenever it was possible for both the incoming
and outgoing channels. 
With these requirements in mind, the set of reactions
used to study the uniqueness property are shown in table (\ref{reactions}).

A spherical two-body model is used 
for describing the ground state of each (B = A + p) system.
We take a Woods-Saxon potential with radius of 1.25 fm, diffuseness a=0.65 fm,
and depth adjusted to give the appropriate binding energy
($\epsilon$(p + $^9$B) = 6.5858 MeV,
 $\epsilon$(p + $^{12}$C) = 1.93435 MeV,  and $\epsilon$(p + d) = 5.4935 MeV).
A spin-orbit term with the same geometry parameters and
depth of 2.06 MeV was included.
The $^{10}$B g.s. was then described as $p_{3/2}$ proton coupled to the
$^9$Be(3/2$^-$) core  ($p_{3/2} \otimes 3/2^-$),
the $^{13}$N g.s. as ($p_{1/2} \otimes 0^+$) and the
$^{3}$He g.s. as ($s_{1/2} \otimes 1^+$).
The two-body $p-^{9}$Be s.p. model is not expected
to provide a complete description for the $^{10}$B system.
In fact, the core ground state is close to the $\alpha \alpha n$ threshold
and other terms  may contribute significantly to the wave function.
Within the present reaction mechanism framework,
the incompleteness of the two-body model in describing  the composite nucleus
is taken into account through the extracted spectroscopic factor
${\cal S}_{\rm exp}$. 

In order to extract the ANC  $C_{19}^2$ from  reaction (${\cal A}$),
we proceed in the same way as in \cite{Akram}. 
The transfer reaction (${\cal B}$) provides similarly information
on the $C^2_{19}$.
The $^{10}{\rm B(d,}^3{\rm He)}^9{\rm Be}$ reaction (${\cal C}$)
can be expressed in terms of the product 
$C^2_{19} C^2_{12}$ where $C_{12}$ is the ANC for the vertex
$^3{\rm He} \rightarrow d + p$ given in \cite{Akram2}.
As for reaction (${\cal D}$), the DWBA cross section can be expressed in 
terms of the
product $ C^2_{19}   C^2_{1~12}  $ where $C_{1~12}$ is the
ANC constant for the transition $^{13}$N $\rightarrow$ p +  $^{12}$C,
that was  extracted from the transfer reaction (${\cal E}$).
In all cases the calculations were performed  using FRESCO \cite{Fresco}.

\section{Results}

\vspace{1 cm}
\noindent
{\large \bf The experimental  analysis}    
\vspace{1 cm}

\begin{figure}[h!]
\centerline{
	\parbox[t]{4.0in}{
	\psfig{file=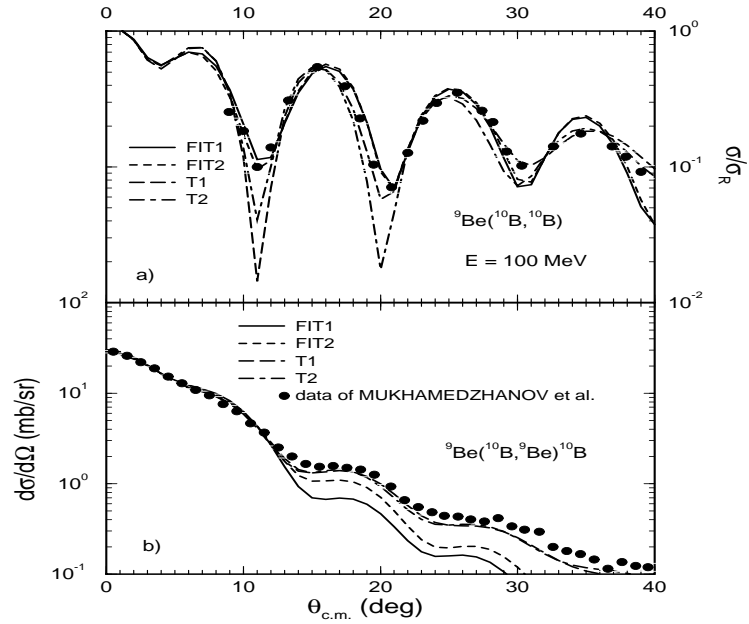,height=3.8in,width=3.8in}}
}
	\caption{Incoming elastic and transfer differential cross section for
$^9$Be($^{10}$B,$^9$Be)$^{10}$B at E$_{\rm lab}$ = 100 MeV. }
	\label{fig:be9b10}
\end{figure}

\begin{figure}[h!]
\centerline{
	\parbox[t]{4.0in}{
	\psfig{file=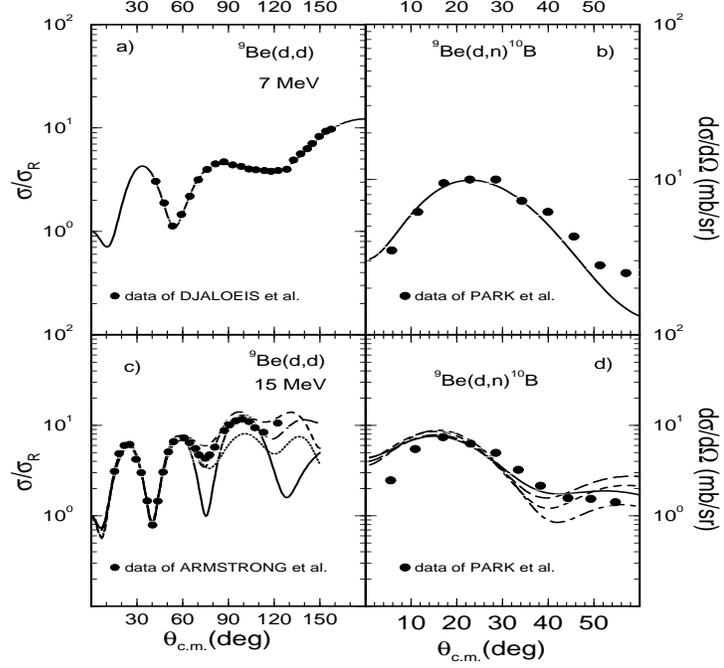,height=3.8in,width=3.8in}}
}
	\caption{Incoming elastic and transfer differential cross sections for
$^9$Be(d,n)$^{10}$B at E$_{\rm lab}$ = 7 and 15  MeV.}
	\label{fig:be9dn}
\end{figure}

\begin{figure}[h!]
\centerline{
	\parbox[t]{4.0in}{
	\psfig{file=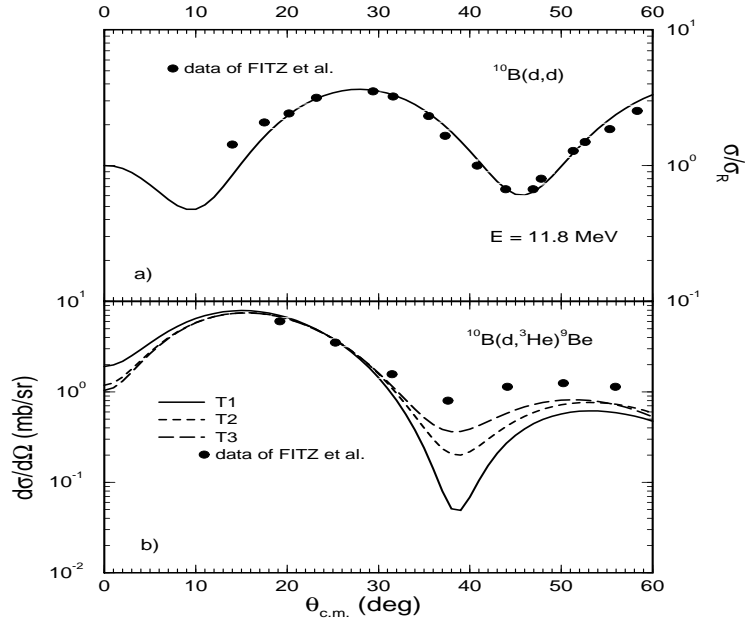,height=3.8in,width=3.8in}}
}
	\caption{Incoming elastic and transfer differential cross sections for
$^{10}$B(d,$^3$He)$^9$Be at  E$_{\rm lab}$ = 11.8 MeV.  }
	\label{fig:be9he3}
\end{figure}

\begin{figure}[h!]
\centerline{
	\parbox[t]{4.0in}{
	\psfig{file=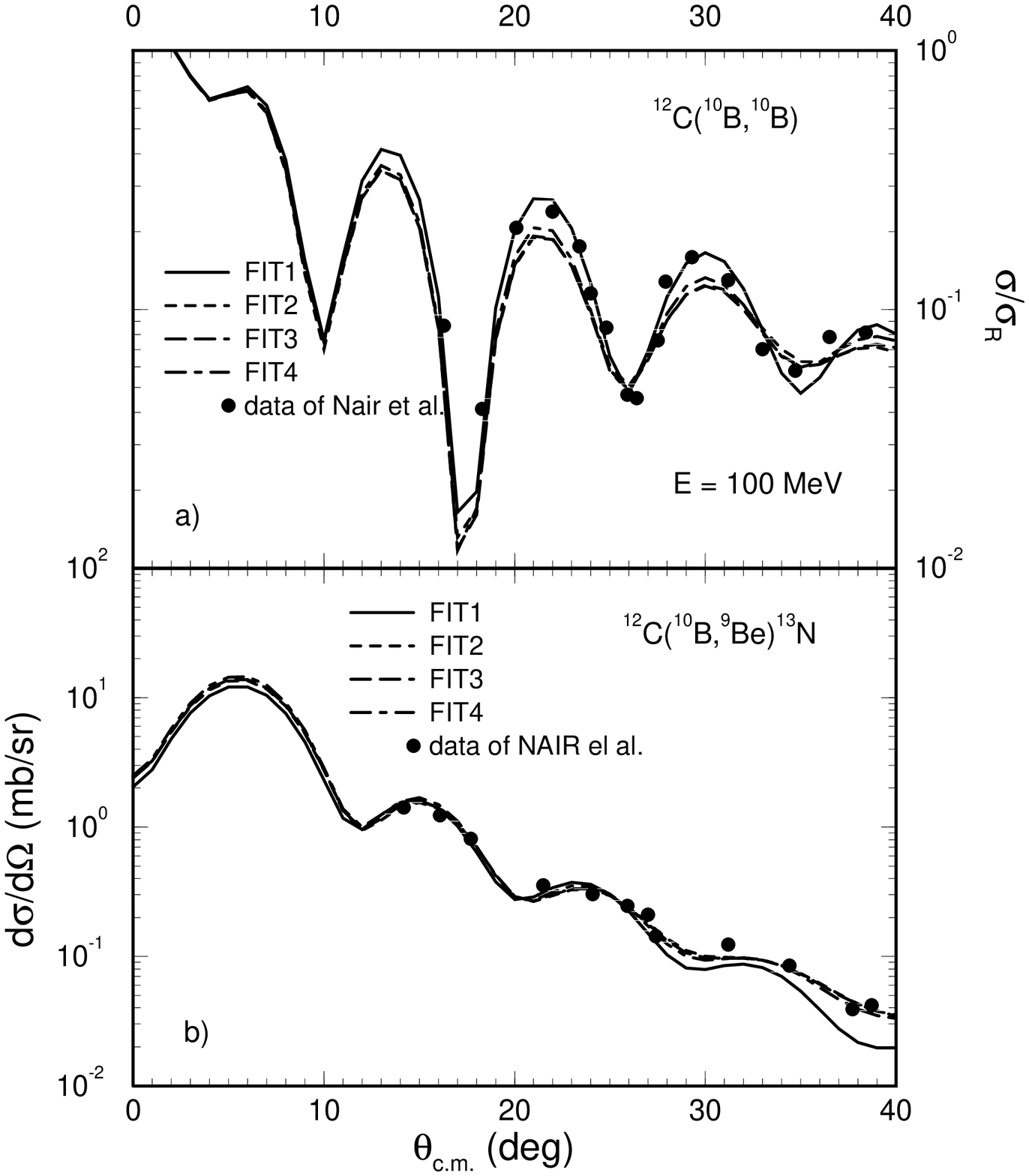,height=3.8in,width=3.8in}}
}
	\caption{Incoming elastic and transfer differential cross sections for
  $^{12}$C($^{10}$B,$^9$Be)$^{13}$N     at E$_{\rm lab}$ = 100  MeV.}
	\label{fig:c12b10}
\end{figure}

\begin{figure}[h!]
\centerline{
	\parbox[t]{4.0in}{
	\psfig{file=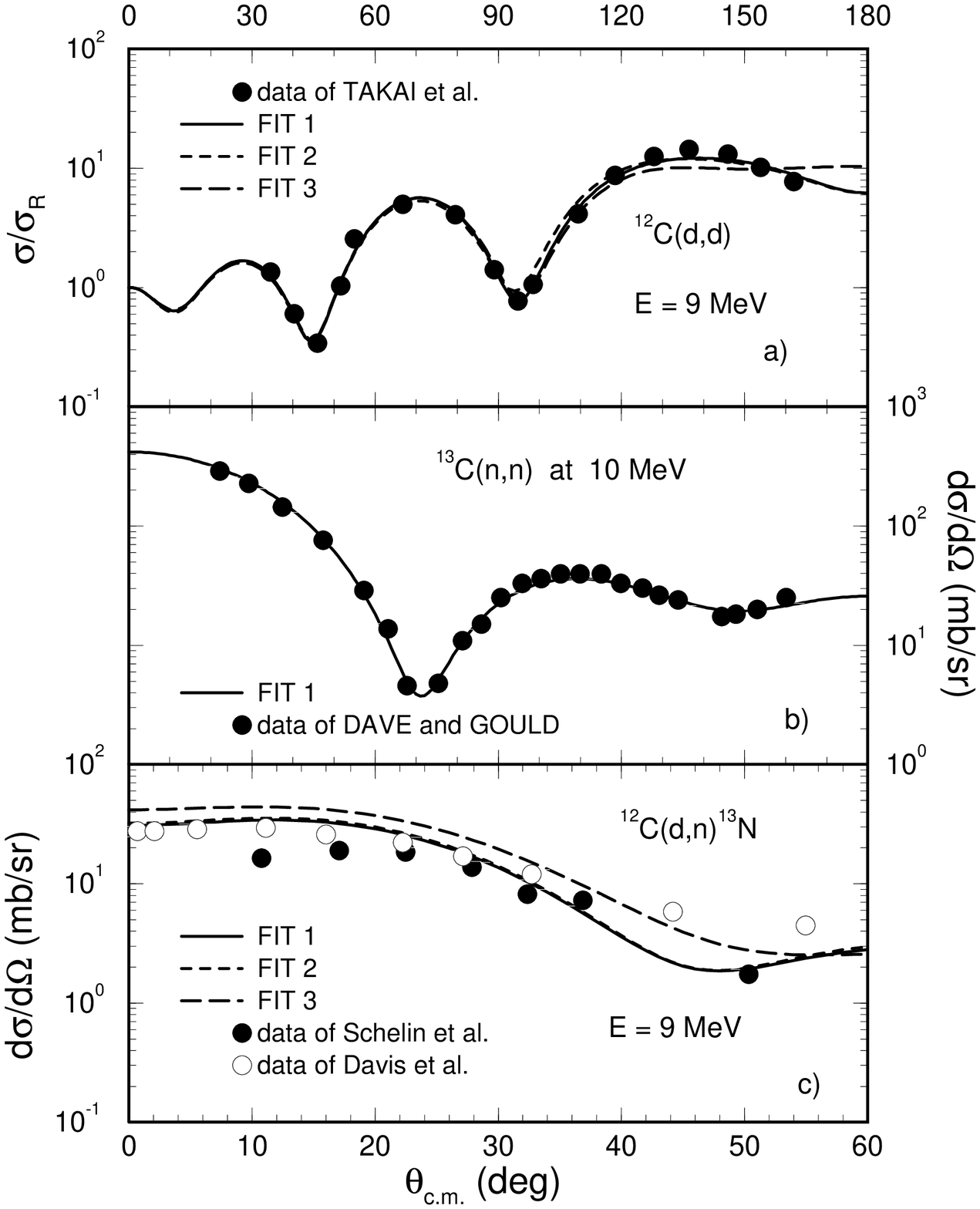,height=3.8in,width=3.8in}}
}
	\caption{Incoming elastic and transfer differential cross sections for
$^{12}$C(d,n)$^{13}$N at E$_{\rm lab}$ = 9  MeV.}
	\label{fig:c12dn}
\end{figure}

\begin{figure}[h!]
\centerline{
	\parbox[t]{4.0in}{
	\psfig{file=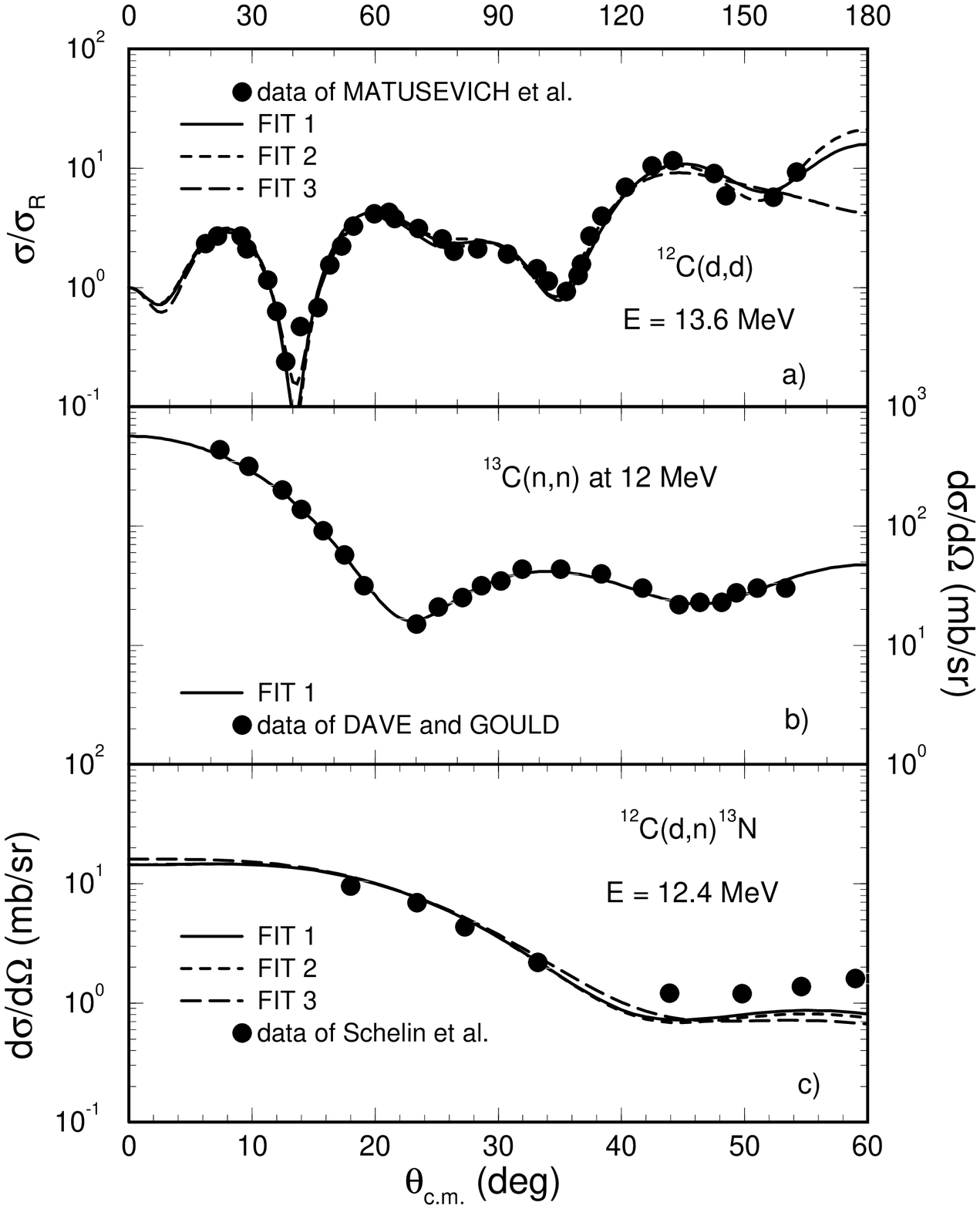,height=3.8in,width=3.8in}}
}
	\caption{Incoming elastic and transfer differential cross sections for
$^{12}$C(d,n)$^{13}$N at E$_{\rm lab}$ = 12.4 MeV.}
	\label{fig:c12dn12}
\end{figure}

The elastic and transfer experimental differential cross sections, for all
the reactions we are considering, are shown in 
figs.(\ref{fig:be9b10}-\ref{fig:c12dn12}).
For each transfer reaction, we firstly determine a set of optical potential
parameters that fit the elastic channels. 
In doing the elastic fit we take into account  
the whole angular range available. The starting parameters for 
this fit were taken from  sets found in the literature at a nearby projectile
energy. These were chosen to have considerable differences in order
to truly evaluate the uncertainties on the ANCs due to  the
choice of optical potentials. 
The parameters for these optical potentials are collected in the  Appendix.

The ANC is directly related to ${\cal S}_{\rm exp}$ (see eq.(\ref{ANC-AC})). 
For a given pair of optical potentials (entrance and exit channels) 
 ${\cal S}_{\rm exp}$ is the normalisation of the forward angle DWBA
cross section  coming from a $\chi^2$ fit to the transfer data. 
The quality of the fit (the accuracy with which the DWBA predicted
angular distribution is able to reproduce the angular dependence
of the data) is quantified by 
$ \chi^2 = \frac{1}{N_{\rm exp}} \sum_i 
\left( \frac{ {\sigma}_{\rm exp}(i) - {\cal S}_{\rm exp} {\sigma}_{\rm Theo}(i) }
{ \Delta {\sigma}_{\rm exp}(i)}\right)^2$ with $N_{\rm exp}$ 
the number of experimental points. These optical potentials are presented in
tables  (\ref{optical1}-\ref{optical4}).
To evaluate the effect of different choices of optical potentials
on the calculated transfer differential cross section,
we also  show in these tables the corresponding 
$\chi^2$ values for the transfer.

Since the aim of this method is to extract an overall normalisation of the
transfer data, it is not only essential to have
data with low statistical errors but, more importantly, low systematic errors.
The uncertainty in the target thickness is a large contributor to the systematic
error, except for the data in \cite{Akram} where special attention was paid to
this issue. For the (d,n) reactionm the neutron efficiency uncertainty is also
quite significant. Other typical errors that may arise are 
due to beam collection or error in solid angle, but are much lower than 
those mentioned above. 
The systematic errors for the set of reactions are collected 
in table (\ref{reactions}), according to the information in the literature. 
It is evident from table (\ref{reactions}) that only the normalisation 
error of the data from \cite{Akram} has the desired low value. 

Another source of error could come from the angular
range from which data is being considered. We must ensure that
the peripheral character of the transfer reaction is satisfied.
A large number of experimental points in the
forward angle region (where the transfer is clearly peripheral)
is desirable, but in most cases non-existent.
For this reason, we have taken a forward angle subset of the data: 
the first 7 points. Even then some sets have angular ranges up to
$\simeq 40 ^{\circ}$ (see table \ref{reactions}).
In two cases a smaller set of data points had to be chosen.
The worst example we have considered is for the (d,n) reactions.

Finally, there will be errors on the derived ANCs
arising from uncertainties on the optical potentials since
the elastic data does not totally probe the interaction. 
The ANC errors shown in table(\ref{S-factors1}) 
and quoted in the text are associated only with the optical potential 
uncertainties.

\begin{figure}[h!]
\centerline{
	\parbox[t]{4.0in}{
	\psfig{file=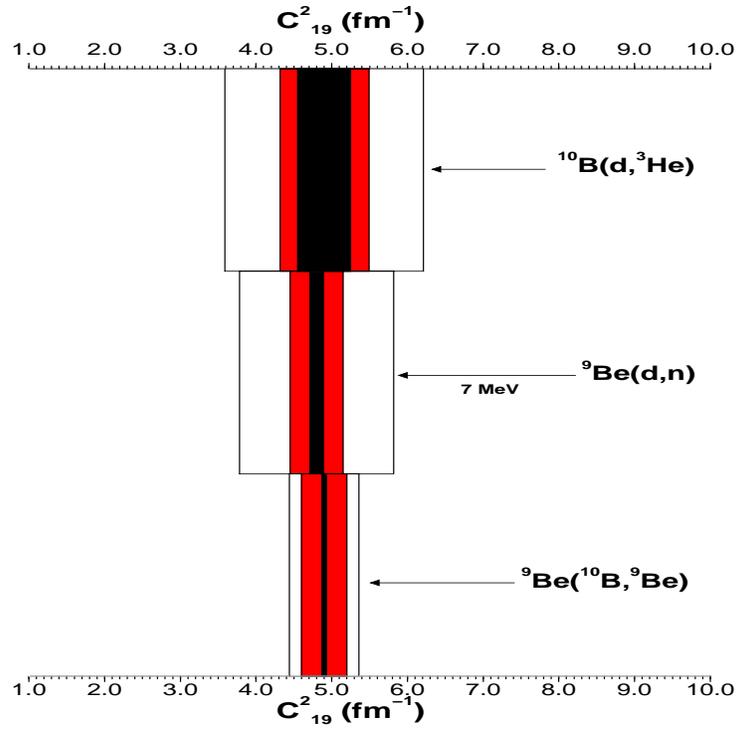,height=3.8in,width=3.8in}}
}
	\caption{Sum of the contributions of the statistical (dark bar),
optical potential (light grey bar) and systematic (white bar) uncertainties
for the calculated ANCs.}
	\label{fig:errors}
\end{figure}

An overall panorama of the errors involved when using this method is given
in fig.(\ref{fig:errors}). 
It is clear from what has been presented in this section that
our results should be interpreted as indicators until further measurements 
are available. They underline the need for further experimental work, 
before definite conclusions on the uniqueness property of the ANC method can
be drawn.

\vspace{1 cm}
\noindent
{\large \bf $^9{\rm Be}(^{10}{\rm B},^9{\rm Be})^{10}{\rm B}$}    
\vspace{1 cm}

The reaction  $^9$Be($^{10}$B,$^9$Be)$^{10}$B is particularly 
adequate for extracting the ANC since it has the same vertex for the
incoming and outgoing channels. The experimental spectroscopic 
factor ${\cal S}_{\rm exp}$ should then be proportional to $C^4_{19}$. 
The elastic and transfer data  at E$_{\rm lab}$=100 MeV was taken
from  \cite{Akram}. 
We take four sets of Wood-Saxon optical model potentials for the incoming
 elastic
channel: the first two obtained by a fitting procedure to elastic data 
shown in fig.(\ref{fig:be9b10}a)
and the others taken from Mukhamedzhanov {\em et al.} \cite{Akram}. 
For this particular reaction, the starting parameter set used in fitting
the elastic data was taken from Comer \cite{Comer} at 40 MeV. The
fits to the  elastic scattering are shown in fig.(\ref{fig:be9b10}a).
The calculated DWBA transfer cross section
renormalised by the spectroscopic factor, fig.(\ref{fig:be9b10}b), 
reproduces quite well the transfer data at small angles.

We note that, the calculated ANC is not strongly dependent on the
details of the optical potentials pinned by the elastic data.
Even if unrealistically shallow potentials are used, the ANC hardly changes. 
Thus, in this case, the uncertainties associated with the choice of
the optical potentials are very small.

\begin{figure}[h!]
\centerline{
	\parbox[t]{4.0in}{
	\psfig{file=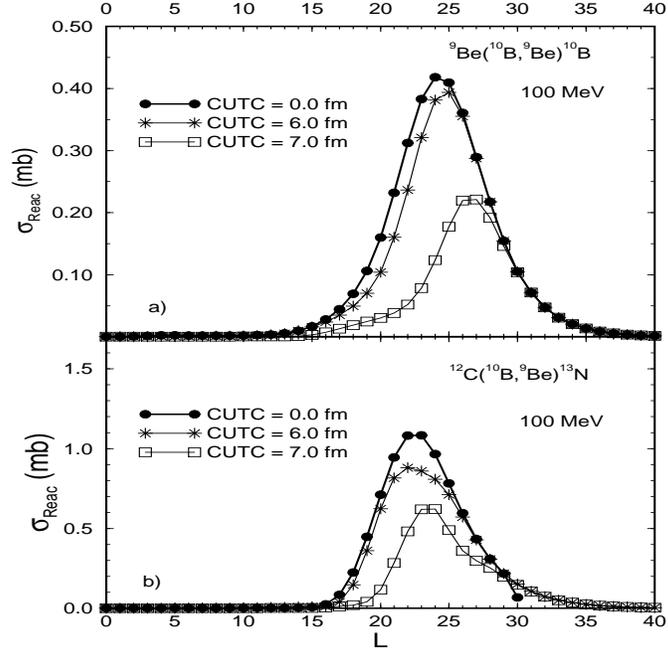,height=3.8in,width=3.8in}}
}
	\caption{ Calculated reaction cross section for each  partial wave L,
for  $^9$Be($^{10}$B,$^9$Be)$^{10}$B  and  $^{12}$C($^{10}$B,$^9$Be)$^{13}$N
at E$_{\rm lab}$ = 100 MeV.   }
	\label{fig:b10be9-cut}
\end{figure}

The calculated reaction cross section as a function of the partial
wave,  fig.(\ref{fig:b10be9-cut}a), clearly shows that this reaction
is peripheral.
In fact, the cross section at E$_{\rm lab}$=100 MeV peaks around L=24
corresponding to an impact parameter of 7.29 fm which is significantly
greater than the sum of the $^9$Be  and $^{10}$B  interaction radius
\cite{radiusRoger,radiusTanihata}.
For that reason the reaction cross section
only becomes significantly smaller for a cuttoff radius much bigger than 
the interaction radius.
We obtained for the ANC, \underline{$C^2_{19}$= 4.9 $\pm$ 0.25  fm$^{-1}$}, 
where, as mentioned before, the error is associated with an optical potential
uncertainty. This is in good  agreement with the value obtained in \cite{Akram}.

\vspace{1 cm}
\noindent
{\large \bf $^9{\rm Be}({\rm d},{\rm n})^{10}{\rm B}$  }
\vspace{1 cm}

The reaction  $^9$Be(d,n)$^{10}$B  provides direct information on
the $C^2_{19}$ as the vertex for the deuteron is well known. 
We performed calculations for 2 different deuteron laboratory
energies: 7 and 15  MeV.

At 7 MeV we used transfer  data from Park \cite{Park} and 
elastic data for $^9$Be(d,d) at 6.3 MeV from Djaloeis \cite{Dja}. 
For the incoming channel, optical model potential parameters were obtained 
from fitting the data shown in  fig.(\ref{fig:be9dn} a).
For the outgoing channel, the potential parameters were 
taken from Dave and Gould \cite{jc8}. 
As follows from the fig.(\ref{fig:be9dn} b), the calculated transfer
differential cross section describes quite well the
 data. We extracted an ANC of 
\underline{$C^2_{19}$ = 4.8 $\pm$ 0.35 fm$^{-1}$} which is in 
good agreement with our previous result obtained from the analysis of
reaction ${\cal A}$. 

\begin{figure}[h!]
\centerline{
	\parbox[t]{4.0in}{
	\psfig{file=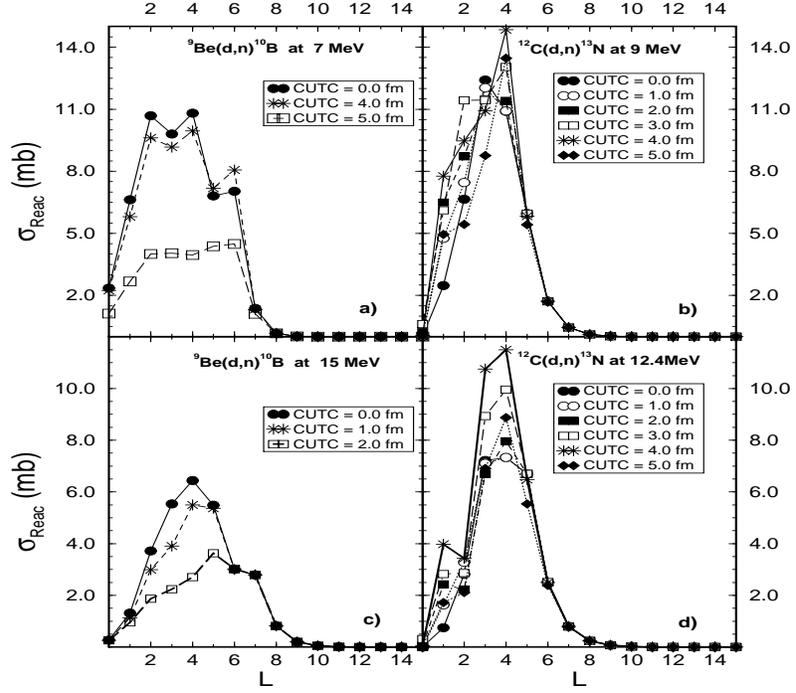,height=3.8in,width=3.8in}}
}
	\caption{    Calculated reaction cross section for each  partial wave L,
for $^{9}$Be(d,n)$^{10}$B at 7 and 15 MeV,
 and  $^{12}$C(d,n)$^{13}$N reactions at 9 and 12.4 MeV.     }
	\label{fig:dn-cut}
\end{figure}

At 15 MeV we also used transfer data from Park $\cite{Park}$ and elastic 
$^9$Be(d,d) at 15 MeV from Armstrong \cite{Arm} published by \cite{Park}. 
Four entrance potential parameter sets  were obtained
fitting the data.
The outgoing parameter set was taken from \cite{jc8} at 
the appropriate energy. 
As can be seen from fig.(\ref{fig:be9dn} c), and fig.(\ref{fig:be9dn} d) 
the 4 parameter sets used
describe quite well the elastic data, but none is able to reproduce 
satisfactory the transfer data in the low angle region
($\theta \leq 20^\circ$). As a result, the calculated ANCs depend 
crucially  on both the input parameter set type, surface or volume 
(about 20 \%), and on the low angle region chosen to minimise 
$\chi^2$ in order to obtain ${\cal S}_{\rm exp}$
({\em vide} table(\ref{S-factors1})).
The derived value for the ANC at this energy is 
\underline{$C^2_{19}$ = 6.09 $\pm$ 0.54 fm$^{-1}$}
 which is  higher than 
that found by \cite{Akram}.
As can be concluded from   fig.(\ref{fig:dn-cut}a) and fig.(\ref{fig:dn-cut}c),
while the reaction is peripheral at 7 MeV,
this is no longer the case for 15 MeV. Thus, this data is not useful for
the purpose of this work.

\vspace{1 cm}
\noindent
{\large \bf $^{10}{\rm B}({\rm d},^3{\rm He})^9{\rm Be}$  }
\vspace{1 cm}

The proton pickup reaction  $^{10}$B(d,$^3$He)$^9$Be
has two different  vertices, $^{10}$B $\rightarrow ^9$Be + p  
and $^{3}$He $\rightarrow$ d + p, and therefore 
the experimental spectroscopic factor 
 will be proportional to the  ANCs product 
$C^2_{19}  C^2_{12} = {\cal S}_{\rm exp} b^2_{19}  b^2_{12}$.

The transfer and elastic data for this reaction at
E$_{\rm lab}$=11.8 MeV was taken from \cite{Fitz}.
For the  entrance  channel, one parameter set was obtained fitting the elastic 
data shown in  fig.(\ref{fig:be9he3}a). 
For the exit channel, we used three parameter sets from literature: 
the first from \cite{Fitz}, the second from \cite{Crosby} and the third from 
\cite{Park}. 
The description of transfer data shown in fig.(\ref{fig:be9he3}b)
is very reasonable, specially for the low angle 
region. 
The analysis of the reaction cross section as a function of the partial
wave number  shows that this transfer reaction is peripheral
fig.(\ref{fig:b10be9-cut}b).
By renormalising the calculated DWBA differential
cross section from the data we obtained 
{$C^2_{19}  C^2_{12} = 19.17 \pm 1.82$ fm$^{-2}$}.
For the $^{3}$He $\rightarrow$ d + p vertex,  we used the value taken 
from \cite{Akram2} $C^2_{12}$ = 3.9 $\pm$ 0.06 fm$^{-1}$. 
Consequently, we get \underline{$C^2_{19}$ = 4.92 $\pm$ 0.54 fm$^{-1}$},
in good agreement with the result of \cite{Akram}.

\vspace{1 cm}
\noindent
{\large \bf $^{12}{\rm C}(^{10}{\rm B},^9{\rm Be})^{13}{\rm N}$ }
\vspace{1 cm}

We proceed in our systematics by looking at other proton stripping reactions
involving $^9$Be. 
A candidate for which measured data was found is
$^{12}$C($^{10}$B,$^9$Be)$^{13}$N at 100 MeV  \cite{Nair}. 
The elastic scattering  fig.(\ref{fig:c12b10}a)
was  taken from  \cite{Nair}. 
We also take the same data for the outgoing channel due to the absence of
experimental measurements for this channel.
When fitting the elastic data for the incoming channel, we
obtain  fouor parameter sets. The optical potential parameters
for the outgoing elastic channel were taken to be the same but with
an appropriate radius as discussed in the Appendix. 
The calculated DWBA cross section describes quite well the 
transfer data  shown in  fig.(\ref{fig:c12b10}b)
specially for sets 2, 3 and 4. 

As this reaction has two different vertices for the two composite nuclei, 
$^{10}{\rm B} \rightarrow ^9{\rm Be} + {\rm p}$ 
and $^{13}$N $\rightarrow ^{12}$C + p, the experimental
 spectroscopic factor 
${\cal S}_{\rm exp}$ will be proportional to the ANCs product, 
$C^2_{19}  C^2_{1~12} = {\cal S}_{exp} \; b^2_{19} \;  b^2_{1~12}$.
We obtained $C^2_{19} C^2_{1~12} = 7.4 \pm 0.5 \: {\rm fm}^{-2}$.

\vspace{1 cm}
\noindent
{\large \bf $ ^{12}{\rm C}({\rm d},{\rm n})^{13}{\rm N}$  }
\vspace{1 cm}

In order to extract $C^2_{19}$ from  the results obtained with the 
last reaction,  it is necessary to extract $C^2_{1~12}$ from 
another independent reaction. We chose 
the $^{12}$C(d,n)$^{13}$N reaction at two different energies: 9 and 12.4 MeV.

At 9 MeV we used transfer data available from two different sources, 
Davis {\em et al.} \cite{Davis} and  Schelin {\em et al.} \cite{Schelin}.
We take 3 sets of potential parameters  from  fitting  the entrance channel
elastic data of \cite{Takai} shown in fig.(\ref{fig:c12dn}a), and one set 
for the exit channel data of \cite{jc8} for $^{13}$C(n,n) at 10 MeV shown in 
fig.(\ref{fig:c12dn}b). 
As shown in fig.(\ref{fig:c12dn}c) the shape of the calculated DWBA
differential cross section  adjusts much better to Davis's data  than to
Schelin's  in the low angle region, $\theta \leq 35^\circ $.
Not surprisingly the derived spectroscopic factors from the two data sets
differ by 30 $\%$. We found
$C^2_{1 ~12} = 2.56 \pm 0.37 $ for  \cite{Schelin} and
$C^2_{1 ~12} = 3.31 \pm 0.45 $ for  \cite{Davis}.
Again, these errors are associated only with the optical potential
uncertainty. 
Using the results for the ANC product from reaction 
 $^{12}$C($^{10}$B,$^9$Be)$^{13}$N we obtain respectively 
\underline{$C^2_{19}$ = 2.98 $\pm$  0.63  fm$^{-1}$}  and
\underline{$C^2_{19}$ = 2.30 $\pm$ 0.47 fm$^{-1}$}.

An energy average  differential cross section data, 
at 12.4 MeV, is given in Schelin's work  $\cite{Schelin}$  
(using 13.0 MeV data  and 11.8 MeV data of Mutchler $\cite{Mutc}$). 
For the entrance channel we used the Matusevich \cite{Matu} experimental points  
at 13.6 MeV  fig.(\ref{fig:c12dn12}a). The exit 
channel data was taken from \cite{jc8} for $^{13}$C(n,n) at 12 MeV 
fig.(\ref{fig:c12dn12}b). Three 
parameter set fits were obtained for the deuteron potential and one for the
neutron potential.

As follows from fig.(\ref{fig:c12dn12}c), a good agreement in the low angle
region ($\theta \leq 40^\circ $) is achieved 
between the  data and the calculated cross section. The calculated 
spectroscopic factors 
lead us to $C^2_{1~12}$ = 1.65 $\pm$ 0.2 fm$^{-1}$.
Using the ANC product from the reaction 
for $^{12}$C($^{10}$B,$^9$Be)$^{13}$N,  one obtains 
\underline{$C^2_{19}$ = 4.6 $\pm$ 0.9 fm$^{-1}$}.  

We note from figures (\ref{fig:dn-cut}b) and (\ref{fig:dn-cut}d)
that while the reaction $ ^{12}{\rm C}({\rm d},{\rm n})^{13}{\rm N}$
is peripheral at 9 MeV, the situation is rather unclear at 12.4 MeV.
Although the impact parameter is greater than the $^{12}$C
interaction radius \cite{radiusRoger}, the reaction cross
section drops significantly for a cutoff radius R$_{\rm cut}$ = 1 fm.

The experimental situation concerning this reaction is rather unsatisfactory
due to different available  data at 9 MeV.
In the works of Davis \cite{Davis} and Shelin \cite{Schelin}, a contribution
due to compound nucleus formation to the cross section is estimated using 
the Hauser-Feshbach statistical model. For the  
$ ^{12}{\rm C}({\rm d},{\rm n})^{13}{\rm N}$ reaction
at 12.4 MeV, the calculated compound nucleus cross section
lead to an overestimation of the cross section at large 
angles. The Hauser-Feshbach model is then clearly inadequate in this case. 
Arbitrary reduction factors  can be found in \cite{Davis,Schelin} 
producing a large uncertainty on the 
derived spectroscopic factors and  ANCs. 
In order to extract a meaningful ANC factor from a transfer reaction, 
the reaction mechanism should be properly understood. 

Since no information on the uncertainty of the absolute cross section
for reaction $\cal D$ is given \cite{Nair} the C$_{19}$ thus extracted
should not be used to validate the ANC method.
Agravating the situation are the differences between the $^{12}$C(d,n) 
data sets suggesting that the C$_{1~12}$ are not sufficiently reliable 
to be taken into account.

\section{Conclusions}

We have determined the ANC  
for the $^{10}$B $ \rightarrow  ^9$Be + p
using a set of  proton-transfer reactions at different energies.
The calculated ANCs from the different reactions 
reproduced in fig.(\ref{fig:errors})  clearly reveals the present experimental
situation if one wants to check the validity of the ANC method. 
The sum of the contributions of the statistical,
optical potential and systematic  uncertainties
is in most cases  quite large.
The graph evidently shows that, from a particular set of transfer reactions 
(those that are clearly peripheral and have quotable normalisation errors),
the uniqueness property of the ANC is satisfied.
However, with the present  data, we cannot undoubtedly
conclude if this property is fulfilled.

More  data for both, the transfer and elastic channels,
with good resolution and carefully normalised cross sections
in the forward angular region, is {\bf crucial}
if we want to unambiguously check  the uniqueness of the ANCs. 
The elastic data is essential to reduce the optical parameter uncertainties.
In the measurements special attention should be paid to minimise the 
uncertainty on the target thickness. For (d,n) reaction it is important
to reduce as much as possible the neutron efficiency error, given that
this may be a large source of uncertainty. 

Even though the DWBA method is widely used, care should be taken  
to fully understand the reaction mechanisms before extracting the ANCs.
Early studies on the deuteron breakup effects on the differential cross
section  indicate  \cite{japanese} that DWBA analysis
may not be  a useful tool to study deuteron transfer reactions. However,
even nowadays, these reactions are still used to extract ANCs \cite{Liu}.
More data on deuteron transfer reactions is necessary in order to have a 
better understanding of the  mechanisms and to check if they can be used
to extract the ANCs.
Generally, further tests on the ANC method, focusing
on the reaction mechanism, should also be performed.

\acknowledgements
This work was supported by Funda\c c\~ao de Ci\^encia
e Tecnologia (Portugal) through grant Praxis PCEX/C/FIS/4/96.
We would like to thank L. Trache for providing  us with the experimental
results for the $^9$Be($^{10}$B, $^9$Be)$^{10}$B reaction and elastic data.

\vspace{3 cm}

\section*{Appendix}

We collect  in tables (\ref{optical1}-\ref{optical4}) the optical potential
parameters obtained by fitting the elastic channels.
The potentials are calculated using the following expressions: 
\be 
{\rm Real ~central:} \, \,
 &U_R& = - \,V \frac{f(r)}{1 + f(r)}~~, \nonumber \\
{\rm Imaginary ~central ~volume:} \, \,
 &U_I& = -\, W \frac{f(r)}{1 + f(r)}~~,  \nonumber \\ 
{\rm Imaginary~ central~ surface:}\, \,
 &U_W& = - \,4 W_d \frac{f(r)}{(1 + f(r))^2}~~, \nonumber \\
{\rm Spin-Orbit:} \, \,
 &U_{SO}& = - \,\frac{4}{r a_{SO}} V_{SO} 
         \frac{f(r)}{(1 + f(r))^2} \,\, \vec{l}.\vec{s}~~,
\ee
with $f(r) = e^{- \frac{r - R}{a_i}} $
and $R = r_i A_T^{{1}/{3}}$  except the $^{12}$C($^{10}$B,$^{9}$Be)$^{13}$N 
case where $R = r_i (A_P^{{1}/{3}} + A_T^{{1}/{3}})$.
For all set  of  optical potentials, we use $r_i = r_0$ for real
central, etc. \\

\begin{table}
\caption{Set of reactions used in our analysis.}
\begin{tabular}{cccccccc}
   A(a,b)B  & E (MeV) & l
   Label & Ref. & Sys. Error
 &  N$_{\rm exp}$ &   $\theta_{\rm min}$ & $\theta_{\rm max}$     \\ \hline
$ ^9{\rm Be}(^{10}{\rm B},^9{\rm Be})^{10}{\rm B}$ 
& 100     & ${\cal A}$   &  \cite{Akram}  & 7  $\%$ & 7 & 0.5$^\circ$
 & 6.5$^\circ$  \\ \hline
$ ^9{\rm Be}({\rm d},{\rm n})^{10}{\rm B} $ 
& 7    &   ${\cal B}$ &  \cite{Park}  & 20  $\%$ & 7 &
  5.7$^\circ$ & 40.$^\circ$  \\ 
& 15  &              &  \cite{Park} & 20 $\%$  & 7 & 10.9$^\circ$ &
 44.3$^\circ$ \\ \hline
$^{10}$B(d,$^3$He)$^9$Be    
& 11.8    &   ${\cal C}$ &  \cite{Fitz}  & 25  $\%$  & 3 &  
19.2$^\circ$ & 31.5$^\circ$ 
 \\ \hline 
$^{12}{\rm C}(^{10}{\rm B},^9{\rm B})^{13}{\rm N}$ 
& 100     &   ${\cal D}$ &   \cite{Nair}  & -  & 7 &  14.2$^\circ$ &
  27.0$^\circ$  \\ \hline
$^{12}{\rm C}({\rm d},{\rm n})^{13}{\rm N}$ 
& 9      &   ${\cal E}$ &  \cite{Schelin}   & 12  $\%$ &  6 &
 10.8$^\circ$ &  36.9$^\circ$ \\
& 9      &   &  \cite{Davis}     & 16   $\%$  &   7 &  0.7$^\circ$
  &  27.1$^\circ$ \\
& 12.4   &   &  \cite{Schelin}   & 20  $\%$  &   4 &  18.0$^\circ$ & 
 33.2$^\circ$ \\
 \end{tabular}
\label{reactions}

\vspace{2 cm}

\caption{Deduced ANC factors for $^{10}$B $\rightarrow ^{9}$Be + p. 
The quoted uncertainties arise from optical model analysis of the elastic
channels. }
\begin{tabular}{ccccc}
   A(a,b)B  & E (MeV) & $\overline{S_{exp}}$ & $\overline{\chi^2}$ & 
$\overline{C^2_{19}}$ (fm$^{-1}$)     \\ \hline
$ ^9{\rm Be}(^{10}{\rm B},^9{\rm Be})^{10}{\rm B}$ 
& 100 & 0.52 & 0.80 & 4.90  $\pm$0.25   \\ \hline
$  ^9{\rm Be}({\rm d},{\rm n})^{10}{\rm B}        $ 
 & 7       &  0.42      & 4.29    & 4.80  $\pm$0.35   \\ 
$^{10}$B(d,$^3$He)$^9$Be    
 & 11.8    & 0.48       & 4.07    & 4.92  $\pm$0.54   \\ 
\end{tabular}
\label{S-factors1}

\vspace{2 cm}

\caption{Optical potential parameters 
for the $^{10}$B + $^{12}$C elastic scattering at 100 MeV.}
\begin{tabular}{ccccccccc} 
Set   &  V   & $r_0$  &  a  & W  & $r_{w}$  &  $a_w$  & $r_c$  & $\chi^2$   
  \\
      & {\tiny (MeV)} &  {\tiny (fm)}  &  {\tiny (fm)}
  &  {\tiny (MeV)}  &  {\tiny (fm)}  &  {\tiny (fm)}  &  {\tiny (fm)}  &  
 \\ \hline
FIT1  & 27.62 & 1.105 & 0.802 & 17.17 & 1.179 & 0.531 & 1.03 & 2.18  \\ 
FIT2  & 129.3 & 0.652 & 0.968 & 25.76 & 0.944 & 0.822 & 1.03 & 1.12  \\ 
FIT3  & 215.2 & 0.548 & 0.956 & 30.41 & 0.900 & 0.841 & 1.03 & 1.29  \\ 
FIT4  & 200.9 & 0.588 & 0.935 & 31.02 & 0.944 & 0.768 & 1.03 & 1.80  \\ 
\end{tabular}
\label{optical1}

\vspace{2cm}

\newpage
\caption{Optical potential parameters for the $^3$He + $^9$Be 
elastic scattering at 11.8 MeV.}
\begin{tabular}{c|c|c|c|c|c|c|c|c|c|c|c|c|c|c} 
 Set  &  V   & $r_0$  &  a  &  W & $r_{w}$  &  $a_w$  & $W_d$ & $r_{d}$  & 
 $a_d$  &  $V_{SO}$ & $r_{SO}$  &  $a_{SO}$ & $r_c$  & $\chi^2$   \\
 & {\tiny (MeV)} &  {\tiny (fm)}  &  {\tiny (fm)}  &  
{\tiny (MeV)}  &  {\tiny (fm)}  &  {\tiny (fm)} 
 & {\tiny (MeV)} &  {\tiny (fm)}  &  {\tiny (fm)}
  &  {\tiny (MeV)}  &  {\tiny (fm)}  &  {\tiny (fm)} 
 &  {\tiny (fm)} &   \\ \hline
 T1 \cite{Fitz}
& 149.3 & 1.100 & 0.733 & & & & 7.650 & 1.980 & 0.700 & 5.000 & 1.100 & 0.733 
& 1.40 & 5.98   \\ 
 T2 \cite{Crosby}
& 109.0 & 1.600 & 0.640 & 22.00 & 1.600 & 0.640 & & & & & & & 1.30 & 3.51  \\ 
 T3 \cite{Park}
& 171.0 & 1.200 & 0.510 & & & & 18.00 & 1.200 & 1.990 & 5.500 & 1.200 & 0.510 
& 1.30  & 2.71  \\ 
\end{tabular}
\label{optical2}

\vspace{2cm}

\caption{Optical potential parameters for  d + A  elastic scattering.}
\begin{tabular}{c|c|c|c|c|c|c|c|c|c|c|c|c|c|c|c|c} 
 A & E & Set  &  V   & $r_0$  &  a  &  W & $r_{w}$  &  $a_w$  &
 $W_d$ & $r_{d}$  &  $a_d$  &  $V_{SO}$ & $r_{SO}$  & 
 $a_{SO}$ & $r_c$  & $\chi^2$   \\
  & {\tiny (MeV)} &  &  {\tiny (MeV)} &  {\tiny (fm)}
  &  {\tiny (fm)}  &  {\tiny (MeV)}  &  {\tiny (fm)}  &  {\tiny (fm)} 
 &  {\tiny (MeV)} &  {\tiny (fm)}  &  {\tiny (fm)}  &
  {\tiny (MeV)}  &  {\tiny (fm)}  &  {\tiny (fm)} 
 &  {\tiny (fm)} &   \\ \hline
$^9$Be & 7 & FIT1 & 130.0 & 1.043 & 0.768 &  &  &  & 78.05 & 1.681 & 0.166 & 7.498 
& 1.956 & 0.184 & 1.30 & 4.28 \\ 
$^9$Be & 15 & FIT1 & 58.49 & 1.321 & 0.781 & &  &  & 9.278 & 1.711 & 0.670 & 1.297 
& 1.063 & 0.473 & 1.30 & 2.91 \\ 
$^9$Be & 15 & FIT2 & 82.45 & 1.015 & 0.986 & 34.51 & 1.145 & 0.867 &  &  & & 0.922 
& 1.040 & 1.104 & 1.25 & 8.35 \\
$^9$Be & 15 & FIT3 & 90.53 & 0.841 & 1.019 & 39.36 & 0.669 & 1.060 &  &  & & 0.787 
& 0.687 & 0.920 & 1.25 & 4.90 \\ 
$^9$Be & 15 & FIT4 & 95.67 & 1.655 & 0.551 &  &  & & 46.99 & 1.457 & 0.300 & 3.293 
& 1.294 & 2.442 & 1.25 & 11.8 \\ 
$^{10}$B & 11.8 & FIT1 & 80.55 & 0.924 & 0.972 & 28.89 & 0.853 & 0.732 & & & & 
5.066 & 0.760 & 0.879 & 1.30 &   \\ 
$^{12}$C & 9 & FIT1 & 130.9 & 0.894 & 0.963 &  &  &  & 9.468 & 2.066 & 0.381 & 
4.302 & 1.512 & 0.184 & 1.30 & 0.19 \\ 
$^{12}$C & 9 & FIT2 & 127.8 & 0.941 & 0.940 &  &  &  & 8.851 & 2.028 & 0.397 & 
4.393 & 1.459 & 0.246 & 1.30 & 0.18 \\ 
$^{12}$C & 9 & FIT3 & 121.6 & 0.925 & 0.968 &  &  &  & 10.74 & 1.854 & 0.405 & 
7.525 & 1.993 & 0.497 & 1.30 & 0.08 \\ 
$^{12}$C & 12.4 & FIT1 & 121.4 & 0.891 & 0.872 & & & & 10.20 & 1.856 & 0.517 & 
2.720 & 0.971 & 1.344 & 1.30 & 3.44 \\ 
$^{12}$C & 12.4 & FIT2 & 111.8 & 0.965 & 0.812 & & & & 8.405 & 1.859 & 0.582 & 
2.870 & 0.638 & 1.072 & 1.30 & 3.38 \\ 
$^{12}$C & 12.4 & FIT3 & 118.5 & 0.864 & 1.023 & & & & 11.77 & 1.677 & 0.503 & 
4.880 & 1.943 & 0.439 & 1.30 & 4.72 \\
\end{tabular}
\label{optical3}

\vspace{2cm}

\caption{Optical potential parameters for n + A elastic scattering. }
\begin{tabular}{c|c|c|c|c|c|c|c|c|c|c|c} 
 A & E & Set  &  V   & $r_0$  &  a  & $W_d$ & $r_{d}$  &  $a_d$  &  $V_{SO}$ & 
$r_{SO}$  & $a_{SO}$ \\ 
 & {\tiny (MeV)} &  & {\tiny (MeV)} & {\tiny (fm)}
 & {\tiny (fm)} & {\tiny (MeV)} & {\tiny (fm)} & {\tiny (fm)} 
& {\tiny (MeV)} & {\tiny (fm)} 
& {\tiny (fm)}    \\ \hline
$^{10}$B & 7 & DG \cite{jc8}
 & 44.45 & 1.387 & 0.464 & 8.757 & 1.336 & 0.278 & 5.500 & 1.150 
& 0.500     \\ 
$^{10}$B & 15 & DG \cite{jc8}
& 42.16 & 1.387 & 0.464 & 14.12 & 1.336 & 0.278 & 5.500 & 1.150 
& 0.500    \\
$^{13}$N & 9 & FIT1 & 68.05 & 0.968 & 0.446 & 18.53 & 1.445 & 0.101 & 7.073 & 
0.631 & 0.194 \\ 
$^{13}$N & 12.4 & FIT1 & 50.50 & 1.203 & 0.329 & 3.902 & 0.400 & 0.867 & 7.096 & 
1.444 & 0.353 \\
\end{tabular}
\label{optical4}

\end{table}

\newpage

\begin{figure}[h!]
\centerline{
	\parbox[t]{4.0in}{
	\psfig{file=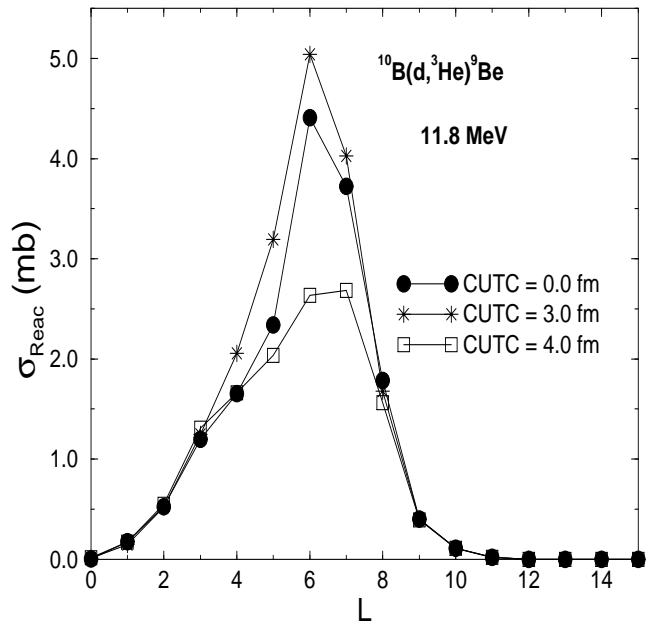,height=3.8in,width=3.8in}}
}
	\caption{ Calculated reaction cross section for each  partial wave L,
for $^{10}$B(d,$^3$He)$^9$Be at  E$_{\rm lab}$ = 11.8 MeV.  }
	\label{fig:he3d-cut}
\end{figure}

\end{document}